\documentclass[%
 reprint,
 amsmath,amssymb,
 aps, longbibliography
]{revtex4-1}

\usepackage{graphicx}% Include figure files
\usepackage{dcolumn}% Align table columns on decimal point
\usepackage{bm}% bold math
\begin{document}

\preprint{APS/123-QED}

\title{Quantifying surface wetting properties \\ using droplet probe AFM}% Force line breaks with \\

\author{Dan Daniel}
 \email{daniel@imre.a-star.edu.sg}
\author{Yunita Florida}
\author{Lay Chee Leng}
\author{Xue Qi Koh}
\author{Anqi Sng}
\author{Nikodem Tomczak}
 \email{tomczakn@imre.a-star.edu.sg}
  
\affiliation{Institute of Materials Research and Engineering, A*STAR (Agency for Science, Technology and Research), 2 Fusionopolis Way, Innovis, Singapore 138634}


\begin{abstract}
Surface wettability has a huge influence on its functional properties. For example, to minimize smudging, surfaces should be able to repel oil droplets. To quantify surface wettability, the most common approach is to measure the contact angles of a liquid droplet on the surface. While well-established and relatively easy to perform, contact angle measurements are crude and imprecise; moreover, they cannot spatially resolve surface heterogeneities that can contribute to surface fouling. To address these shortcomings, we report on using an Atomic Force Microscopy (AFM) technique to quantitatively measure the interaction forces between a micro-droplet and a surface with piconewton force resolution. We show how our technique can be used to spatially map topographical and chemical heterogeneities with micron resolution.
\end{abstract}

\maketitle

%\tableofcontents
There is a huge interest in tuning the wetting properties of surfaces to repel various liquids, improve the rate of heat transfer, and more generally to prevent fouling \cite{quere2008wetting, bocquet2011smooth, attinger2014surface}. To quantify the wetting properties of surfaces, the most common approach is to measure the contact angles of a millimetric-sized liquid droplet \cite{Liu1147, huhtamaki2018surface}. The adhesion and friction forces required to remove the droplet can then be deduced, albeit indirectly, from the advancing and receding contact angles \cite{samuel2011study, furmidge1962studies}. Despite the popularity of such an approach, contact angles do not adequately describe the surface wetting properties \cite{decker1999physics, schellenberger2016water, daniel2018origins}. For example, they cannot capture the local wetting variations due to chemical heterogeneities or surface texture \cite{liimatainen2017mapping}. Moreover, contact angle measurements become inaccurate for large contact angles close to 180$^{\circ}$ \cite{Liu1147, srinivasan2011assessing}.

To overcome the limitations described above, several alternative surface characterization techniques have been proposed \cite{tadmor2009measurement, daniel2017oleoplaning, timonen2013free, liimatainen2017mapping}. More recently, we introduced a modified Atomic Force Microscopy (AFM) method to precisely quantify the wetting properties of superhydrophobic surfaces, by directly measuring the adhesion and friction forces of small water droplets (tens of micron in size) with nN force resolutions \cite{daniel2019mapping}. With this droplet probe AFM technique, we were also able to map local wetting variations due to the presence of micro-/nano-textures and observed fast pinning-depinning dynamics as the droplet detached from the surface. Previously, droplet probe AFM was also used to study various interfacial problems \cite{xie2017surface, shi2016long, manor2008hydrodynamic, escobar2017measuring}, but its significance as a surface characterization tool remains underappreciated. 

In this paper, we show how droplet probe AFM can be used to precisely quantify the wetting properties of various surfaces, such as the water-repellent superhydrophobic surfaces and underwater oil-repellent polyzwitterionic surfaces, and more importantly probe the physical origins of their super-repellent properties. Our technique can also spatially map local wetting properties of surfaces due to both topographical and chemical heterogeneities with piconewton force and micron lateral resolutions. Droplet probe AFM is a versatile, ultra-sensitive technique that will greatly advance wetting science and inform the design of functional coatings. 

\section*{Results and discussions} 

\textbf{The physical origins of liquid-repellent surfaces.} We first quantified (and compared) the wetting properties of a superhydrophobic surface with those of an underwater superoleophobic surface. Specifically, we chose the wings of a Morpho butterfly and glass slides grafted with zwitterionic poly(sulfobetaine methacrylate) (pSBMA) brushes as the two archetypal surfaces. Details on how we prepared the samples have been discussed in our previous publications and elsewhere \cite{daniel2019hydration, daniel2019mapping, azzaroni2006ucst, kobayashi2012wettability}.  

\begin{figure*}[!htb]
\centering
\includegraphics[scale=0.9]{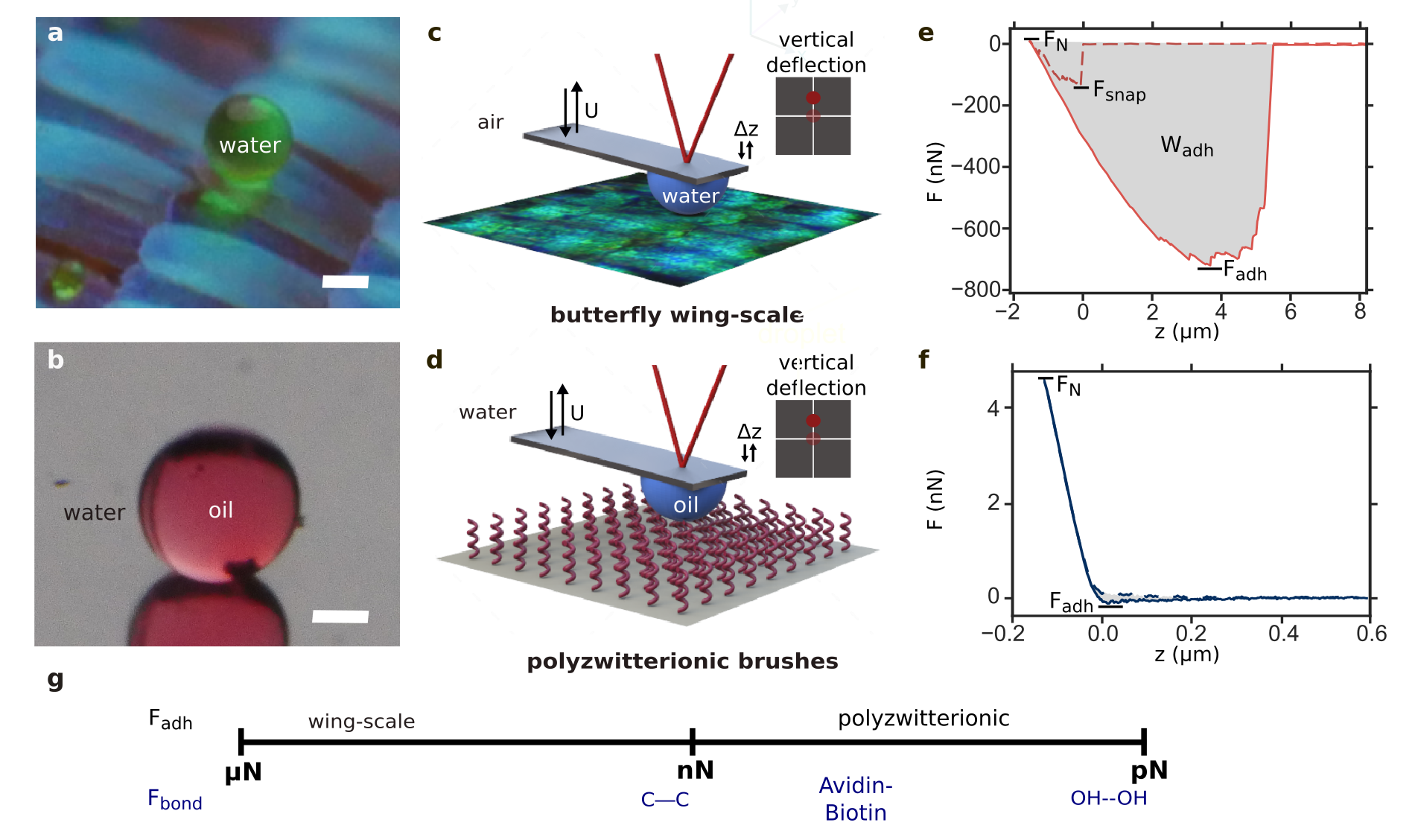}
\caption{\label{fig:origin} (a, b) Water and oil droplets remain spherical on butterfly wing-scales and zwitterionic pSBMA brush surface (under water), respectively. Colorants were added to improve droplet visibility when taking the two photographs. Scale bars are 50 $\mu$m. (c, d) Droplet probe AFM can be used to quantify the repellency of the two surfaces. The cantilever deflection $\Delta z$ was monitored by shining a laser light (infra-red, wavelength of 980 nm) onto the cantilever tip, which was reflected into a four-quadrant sensor. (e, f) Force spectroscopy results for a 30-$\mu$m-sized water droplet and 40-$\mu$m-sized oil droplets moving at $U$=10 $\mu$m s$^{-1}$, respectively. The droplet position $z$ has been corrected for cantilever deflection. (g) Comparison between the adhesion force $F_{\text{adh}}$ required to remove the droplets from the interface and the magnitudes of various chemical bonds.}
\end{figure*}

On both surfaces, the droplets (water and silicone oil, respectively) have high contact angles close to 180$^{\circ}$ (Fig.~\ref{fig:origin}a, b). It is therefore tempting to think of polyzwitterionic surfaces as the underwater oil-repellent analogues of superhydrophobic surfaces \cite{liu2017nature}. However, this analogy is inaccurate and overlooks the different physical origins that give rise to the liquid repellency in the two surfaces. Water-repellency on superhydrophobic surfaces is achieved by the ability of micro-/nano-structures to trap air pockets and minimize (though not eliminate) droplet-solid contact to the topmost tips of the structures. In contrast, polyzwitterionic brushes rely on electric double-layer forces to stabilize a continuous water film beneath the oil droplet. Droplet contact is completely eliminated, resulting in ultra-repellent properties \cite{daniel2019hydration}.
 
\begin{figure*}[!htb]
\centering
\includegraphics[]{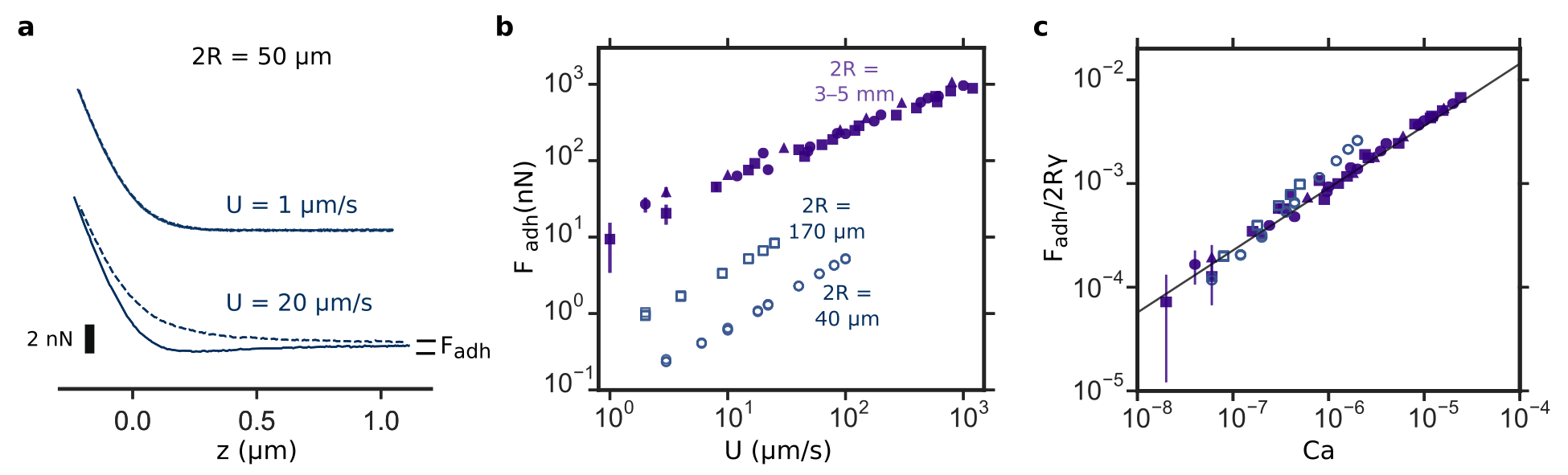}
\caption{\label{fig:psbma} (a) Force spectroscopy results for a 50-$\mu$m-sized oil droplet on the pSBMA brush surface at speeds $U$ = 1 and 20 $\mu$m s$^{-1}$. (b) The adhesion force $F_{\text{adh}}$ depends on the droplet's diameter $2R$ and velocity $U$. Note that $F_{\text{adh}}$ for $2R$ = 3--5 mm were measured using the Droplet Force Apparatus, as reported in \citeauthor{daniel2019hydration} (2019) \cite{daniel2019hydration}. Different droplet sizes are indicated by different filled markers (c) Plot of the non-dimensionalized adhesion force $F_{\text{adh}}/2R\gamma$ against the capillary number $Ca = \eta U \gamma$. Each point respresents a single adhesion force measurement. Error in AFM measurements for sub-millimetric droplets (open markers) are smaller than the marker size. For millimetric droplets (filled markers), the error bars are due to instrument noise, as explained in details in \citeauthor{daniel2019hydration} (2019) \cite{daniel2019hydration}.}
\end{figure*}

To demonstrate the quantitative (and qualitative) differences between the two surfaces, we performed force spectroscopy measurements using droplet probe AFM. Briefly, we attached either a 40 wt$\%$ glycerol-water droplet (in air) or an oil droplet (under water) onto a tipless cantilever probe with a flexular spring constant of $k_z$ = 0.2--2 N m$^{-1}$ (Fig.~\ref{fig:origin}c, d). Glycerol was added to the water droplet to minimize evaporation. The force on the droplet can be deduced from the cantilever deflection $\Delta z$, i.e. $F = k_z \Delta z$, as it approaches and retracts from the surface at a controlled speed $U = 10$ $\mu$m $s^{-1}$. Details on how we obtained $\Delta z$ from the raw voltage signals on the four-quadrant sensor can be found in our previous publication \cite{daniel2019mapping}.   

Figures \ref{fig:origin}e and \ref{fig:origin}f show the force spectroscopy results for the Morpho butterfly wing and the pSBMA surface, respectively. When the 30-$\mu$m-sized water droplet first contacted the butterfly wing, there was a sudden snap-in force $F_{\text{snap}}$ = 132 nN. We continued to press onto the microdroplet to a maximum normal loading force $F_{\text{N}}$ = 10 nN, before retracting (solid line in Figure \ref{fig:origin}e). For the droplet to be completely detached from the surface, there was a maximum adhesion force that had to be overcome $F_{\text{adh}}$ = 720 nN. Detailed analysis of the pinning-depinning dynamics can be found in our previous publication \cite{daniel2019mapping}. 

%, comparable to the force required to break about a hundred covalent carbon-carbon sigma bonds (Fig.~\ref{fig:origin}g) \cite{evans2001probing}

In contrast, on pSBMA surfaces, the charged sulfobetaine groups are able to support strong hydration shells. Moreover, both the polymer brush surface and the oil droplet become negatively charged under water. The electric double-layer forces prevent contact of the oil droplet and therefore eliminate contact-line pinning on pSBMA surfaces \cite{daniel2019hydration}. Hence, there was no $F_{\text{snap}}$ for the 40-$\mu$m-sized oil droplet pressing on the surface, and the retract curve remains smooth and continuous throughout. The adhesion force $F_{\text{adh}}$ = 85 pN is also much smaller, comparable to the force required to break tens of hydrogen bonds (Fig.~\ref{fig:origin}g) \cite{evans2001probing}, despite the comparatively large droplet size (40 $\mu$m) and contact size ($\sim$ 1 $\mu$m).

By integrating under retract curve, we can also obtain the amount of work required to remove the droplets $W_{\text{adh}}$ = 5.6 pJ and 0.013 fJ for the superhydrophobic and polyzwitterionic surfaces, respectively---a fraction of the total surface energy of the droplet $4 \pi R^{2} \gamma \approx$ 200 pJ, where $R$ and $\gamma$ are the droplet's radius and surface tension.

\begin{figure}[!htb]
\centering
\includegraphics[scale=1.0]{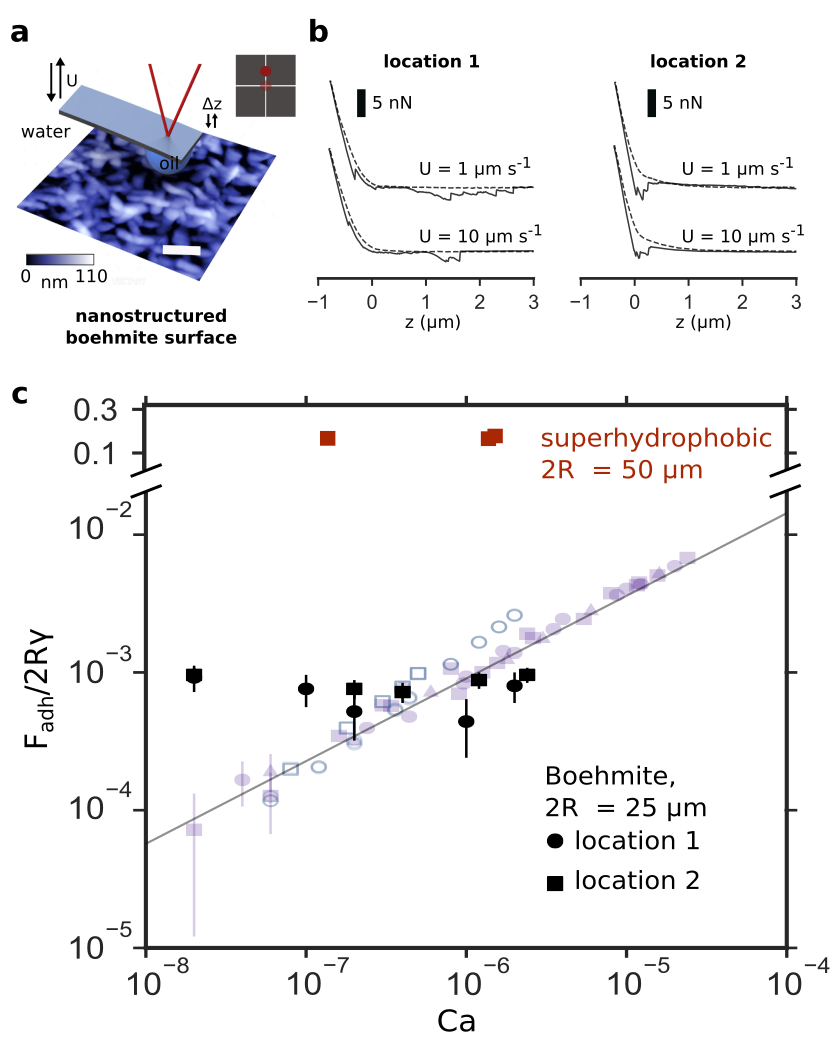}
\caption{\label{fig:boehmite} (a) Underwater superoleophobic properties of nanostructured boehmite surface are quantified using droplet probe AFM. Surface topography was mapped using conventional tapping mode AFM. Scale bar is 200 nm. Droplet and surface are not to scale. (b) Force spectroscopies of 25-$\mu$m-sized oil droplets on two separate locations. (c) Plot of the non-dimensionalized adhesion force $F_{\text{adh}}/2R\gamma$ against the capillary number $Ca = \eta U/ \gamma$ for boehmite surface (black markers). Error bars are standard deviations for 3 repeats. Results for pSBMA surface (blue and purple markers) and superhydrophobic surfaces (red markers, adapted from Figure S11 in \citeauthor{daniel2019mapping} (2019) \cite{daniel2019mapping}) are superimposed on the same plot.}
\end{figure}

The differences between the two surfaces become more striking when we vary the approach and retract speeds $U$. Previously, we have established that for superhydrophobic surfaces $F_{\text{adh}}$ is dominated by contact-line pinning and hence independent of $U$ \cite{daniel2019mapping}. In contrast, for pSBMA surfaces, $F_{\text{adh}}$ is primarily due to viscous forces and hence increases with $U$. At low speed $U = 1$ $\mu$m s$^{-1}$, $F_{\text{adh}}$ is too low to be measured for 50-$\mu$m-sized droplet, but increases to about 2 nN at $U$ = 20 $\mu$m s$^{-1}$ (Fig.~\ref{fig:psbma}a). For 40 and 170-$\mu$m-sized droplets, $F_{\text{adh}}$ even increases to tens and a hundred nN at $U = 0.1$ mm s$^{-1}$ (open circles and squares in Fig.~\ref{fig:psbma}b).    

Previously, we showed that for millimetric oil droplets, $F_{\text{adh}}$ is due to viscous dissipation in the water film beneath the oil droplet (filled markers in Figures \ref{fig:psbma}b, c) and follows the scaling law
\begin{equation} \label{eq:F_adh}
\begin{split}
F_{\text{adh}} \sim 2R \gamma Ca^{3/5},
\end{split}
\end{equation}
 where $Ca = \eta U/\gamma$ is the capillary number, $\eta$ is the water viscosity, and $\gamma$ = 40 mN/m is the oil-water interfacial tensions (line in Figure \ref{fig:psbma}c). $F_{\text{adh}}$ measurements for millimetric droplets were performed using a custom-built Droplet Force Apparatus (DFA). Its working principle is similar to that of an AFM, except with a much more pliant cantilever of $k_z < 0.01$ N m$^{-1}$  \cite{daniel2019hydration, daniel2017oleoplaning}. 
 
 Here, we show that the scaling law remains true for sub-millimetric droplets as measured using droplet probe AFM (open markers in Figures \ref{fig:psbma}b, c). Viscous dissipation in water-film dominate over a wide range of experimental conditions, for droplet sizes between 40 $\mu$m and 5 mm, as well as speeds between 1 $\mu$m s$^{-1}$ and 1 mm s$^{-1}$. Moreover, since the non-dimensionalized $F_{\text{adh}}$ measurements from two instruments (DFA and AFM) overlap with each other, we conclude that our results are not affected by the choice of the cantilevers or measurement systems.

Finally, we would like to point out that not all underwater superoleophobic surfaces are able to eliminate contact-line pinning the way pSBMA surfaces do. Notably, structured oxide surfaces such as boehmite behave more akin to superhydrophobic surfaces, because the oil droplet remains in contact with the topmost tips of the nanostructures (Fig.~\ref{fig:boehmite}a) \cite{liu2009bioinspired, liu2017nature}. Hence, the retract curve is discontinuous with discrete jumps (Fig.~\ref{fig:boehmite}b) and $F_{\text{adh}}$ is insensitive of $U$, just like superhydrophobic surfaces (Fig.~\ref{fig:boehmite}c). Interestingly, boehmite surfaces are more repellent than pSBMA surfaces at high $U$. The choice for the optimal super-repellent surface is therefore dependent on the exact experimental conditions. \\

\begin{figure*}[!htb]
\centering
\includegraphics[scale=1.0]{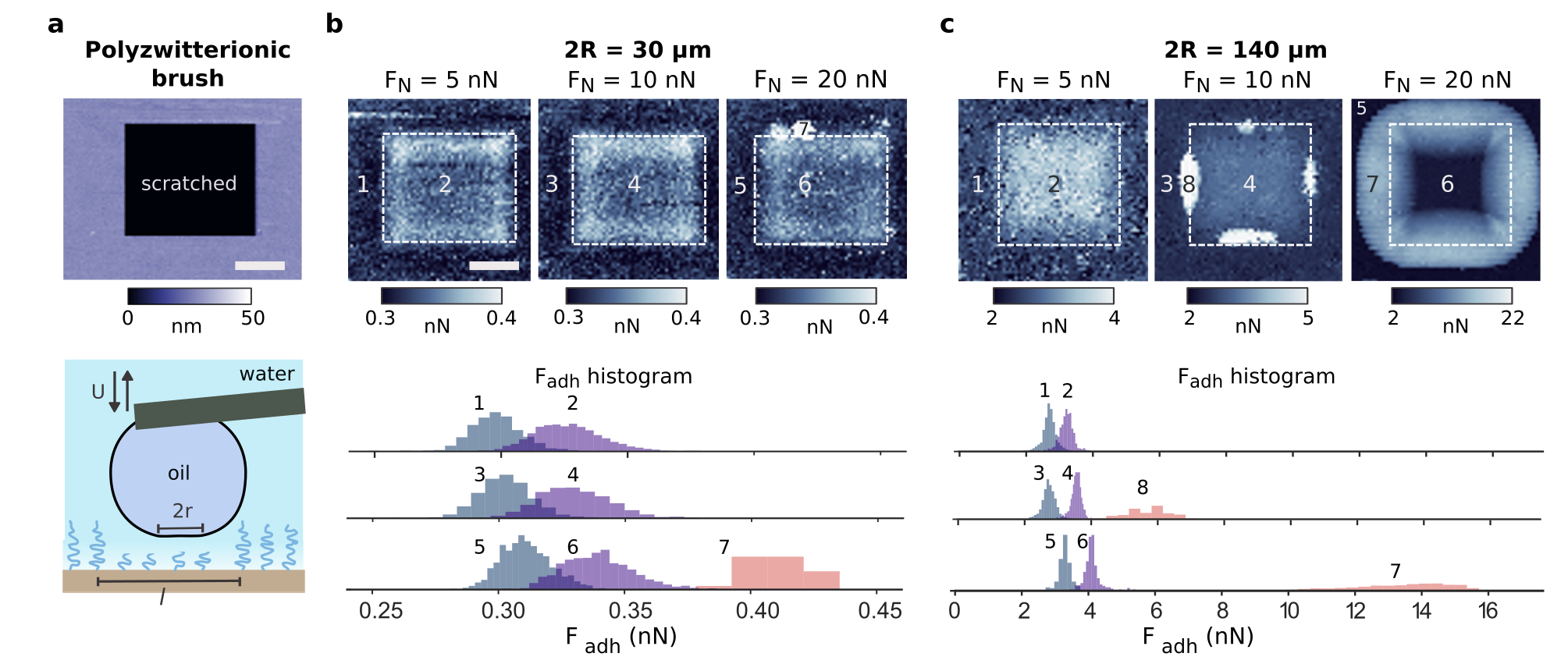}
\caption{\label{fig:topo} (a) 10-$\mu$m-sized square defects were created by scratching the pSBMA brush surface using contact-mode AFM (sharp solid tip) at high load, which were then mapped using Droplet probe AFM. Scale bar is 4 $\mu$m. (b, c) Adhesion force map using oil droplets of diameters $2R$ = 30 and 140 $\mu$m, respectively, at different applied setpoints $F_{\text{N}}$ = 5--20 nN. The Droplet's speed was kept at $U$ = 10 $\mu$m s$^{-1}$. Scale bar is 4 $\mu$m. The histograms were generated from more than 3000 separate measurements.}
\end{figure*}

\begin{figure*}[!htb]
\centering
\includegraphics[scale=0.95]{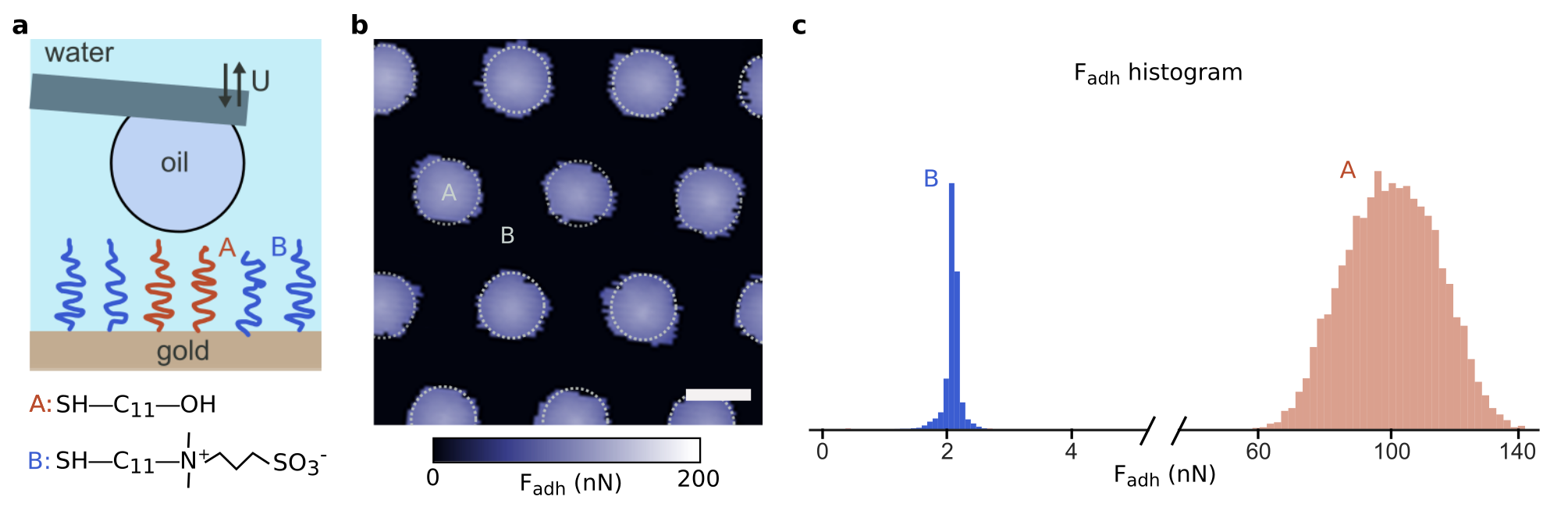}
\caption{\label{fig:thiol} (a) Droplet probe AFM can be used to spatially map micron-scale chemical heterogeneities. (b) Adhesion force map for 5-$\mu$m-sized circular patches of hydroxyl groups (labelled A, dashed lines show the outlines) arranged in a hexagonal array on a gold surface with zwitterionic groups (labelled B). The droplet's size and speed are 50 $\mu$m and 50 $\mu$m s$^{-1}$, respectively. Scale bar is 5 $\mu$m. (c) Histogram for the adhesion force for regions A and B (6258 and 16242 separate measurements, respectively). }
\end{figure*}

\textbf{Mapping topographical and chemical heterogeneities.} We can also exploit the raster scanning capability of the AFM and perform force spectroscopy measurements over a grid array of points to spatially map micron-scale surface wetting variations with pN force resolution. 

To illustrate this, we introduced 10-$\mu$m-sized square defects on the pSBMA surface by performing contact-mode AFM with a conventional sharp-tipped cantilever ($k_z$ = 20 N m$^{-1}$) under a high applied load of 0.2 $\mu$N. This creates sufficient pressure to scratch out the polymer, as shown by the topography map in Figure \ref{fig:topo}a. We then performed droplet probe AFM over the defect to create adhesion force maps for droplets of different sizes $2R$ = 30 and 140 $\mu$m, and under different loading forces $F_{\text{N}}$ = 5--20 nN (Fig.~\ref{fig:topo}b, c). 

Not surprisingly, $F_{\text{adh}}$ is higher where the defects were introduced (regions 2, 4, and 6) compared to the surrounding intact areas (regions 1, 3, and 5). However, the differences in $F_{\text{adh}}$ for the two regions are small (down to several tens of pN for 30-$\mu$m-sized droplet), which indicate that some of the polymer brushes still remain at the scratched areas as shown in the schematic of Fig.~\ref{fig:topo}a.

In contrast, above a threshold $F_{\text{N}} \geq 10$ nN, areas with significantly higher $F_{\text{adh}}$ (regions 7 and 8) started to appear at the edges, presumably because the oil droplet is able to displace the water film at larger $F_{\text{N}}$. See Supplementary Figure S1 for force spectroscopy results for the different regions.  

The adhesion map generated using droplet probe AFM is a convolution of the size of the droplet's base $2r$ and the underlying sample heterogeneities. Since $F_{\text{N}} = (2\gamma/R) \pi r^{2}$ and hence $r \sim (F_{\text{N}} R/\gamma)^{1/2}$, the spatial resolution achievable with the technique decreases with increasing $R$ and $F_{\text{N}}$ (Fig.~\ref{fig:topo}b, c). For 2R = 30--140 $\mu$m and $F_{N}$ = 10--20 nN, this translates to $r \sim 1$ $\mu$m and hence the experimentally observed lateral resolution of about a micron. It should also be possible to achieve sub-micron resolutions with this technique with smaller droplet probes, and this is the subject of our future research direction.     

 Using Droplet Probe AFM, it is also possible to spatially map chemical heterogeneities on the surface. To illustrate this, we patterned a template-stripped gold surface with an array of 5-$\mu$m-sized circular patches of hydroxyl groups (marked A in Fig.~\ref{fig:thiol}) against a background of zwitterionic sulfobetaiene groups (marked B in Fig.~\ref{fig:thiol}) by micro-contact printing \cite{qin2010soft}. The height differences between the two regions are less than a nanometre (Supplementary Fig.~S2) 

The wetting variations due the two different chemical groups can be mapped with micron lateral resolutions, even though the size of the droplet probe at 2$R$ = 50 $\mu$m is much bigger (Fig.~\ref{fig:thiol}b, c). While both hydroxyl and zwitterionic groups are hydrophilic, only zwitterionic groups are able to retain strong hydration shells (due to the presence of electric charges) and eliminate contact. $F_{\text{adh}}$ is therefore much lower for the zwitterionic groups (2.0 $\pm$ 0.5 nN) than the hydroxl groups (100 $\pm$ 15 nN). 

Chemical heterogeneities can be mapped using tapping mode AFM and other chemical force AFM mapping techniques \cite{frisbie1994functional, wong1998covalently, schonherr2005}. In tapping mode AFM with a sharp solid tip, nanometric lateral resolution can easily be achieved and the different chemical groups will manifest themselves as different phases (Supplementary Fig.~S2). It is, however, difficult to interpret the phase values and relate them to wetting properties. In contrast, the physical interpretation for the different $F_{\text{adh}}$ values in droplet probe AFM are clear: the higher the $F_{\text{adh}}$, the more wettable the surface is. 

Finally, we would like to point out that in conventional contact angle measurements, the millimetric droplet size is too large to resolve micron-scale heterogenities as we have done here. This is because the wetting properties probed by contact angle measurements are necessarily averaged over the droplet contact size of about 0.1 mm. The presence (or absence) of heterogeneities, whether topographical or chemical, can have a huge impact on the ability of surfaces to resist fouling \cite{hoek2003effect, mi2010organic}.

\section*{Conclusions}
Since its invention in 1986, Atomic Force Microscopy (AFM) has become a standard and powerful surface characterization tool. Over the years, different variants and imaging modes of AFM have been developed to quantify different physicochemical aspects of  surfaces, from nanomechanical and magnetic to electrical properties. In this context, droplet probe AFM should be seen as complementary to all these imaging modes in providing additional information on the surface wetting properties. Droplet probe AFM is a highly versatile tool that can be used for different liquid probes and surfaces, and if widely adopted, can help resolve many outstanding questions in wetting science.  

\section*{Materials and Methods}

\textbf{Materials} 
\textit{Chemicals for brush growth:} (3-trimethoxysilyl)-propyl 2-bromo-2-methylpropionate (Br-initiator, 95 \%) was purchased from Gelest Inc. [2-(Methacryloyloxy)ethyl]dimethyl-(3-sulfopropyl) ammonium hydroxide (SBMA, $\geq$ 96 \%), copper(I) bromide (CuBr, $\geq$ 98\%), and 2,2\'-bipyridine (bipy, $\geq$ 99 \%) were purchased from Sigma-Aldrich. Methanol was purchased from J.T. Baker. All chemicals were used as received. 

\textit{Probe liquid droplets:} Silicone oil (polyphenyl-methylsiloxane, viscosity $\sim$ 100 mPa.s, density 1.06g/ml) was purchased from Sigma-Aldrich. The water-oil interfacial tension is 40 mN/m for silicone oil as measured using the pendant drop method. Deionized water with a resistivity of 18.2 M$\Omega$.cm was obtained from a Milli-Q water purification system (Millipore, Bedford, MA, USA). 

\textit{Thiolated gold surfaces:} Ultra-flat template-stripped gold surfaces (roughness $<$ 1 nm) were obtained from Platypus technologies. Zwitterionic sulfobetaine-3-undecanethiol was purchased from Dojindo Molecular Technologies, while undecanolthiol was purchased from Sigma-Aldrich. Elastomeric stamp with a hexagonal array of micropillars (5 $\mu$m diameter $\times$ 5 $\mu$m spacing $\times$ 5 $\mu$m tall) was purchased from Research Micro Stamps.  \\
%  
%\textbf{Nanostructured boehmite surface.} 2 nm of Cr film was coated onto clean glass slides, followed by deposition of Al film up to 100 nm with a thermal evaporator system (Syskey, Taiwan) operating under high vacuum of $10^{-6}$--$10^{-7}$ Torr. The coated glass slides were then boiled in DI water close to 100$^{\circ}$ for 20 minutes. This converts the Al coating into nanostructured boehmite layer.\\

\textbf{Zwitterionic PSBMA surface preparation.} The polymer brush surfaces are prepared using surface initiated Atom Transfer Radical Polymerization (ATRP) using a protocol adapted from \citeauthor{azzaroni2006ucst} (2006) \cite{azzaroni2006ucst}.

\textit{Initiator monolayer deposition:} The surfaces (glass or silicon wafer) were rinsed extensively with deionized (DI) water, and then ethanol, before drying under a nitrogen stream. The dried surfaces were then subjected to oxygen plasma surface cleaning at 150 W for 120 s. Br-initiator was vapour deposited onto the cleaned surfaces. In a typical procedure, the cleaned surfaces were heated in a vacuum oven at 75$^{\circ}$C with the Br-initiator (100 $\mu$L) overnight. The silanized surfaces were then cleaned (by rinsing with anhydrous toluene, ethanol, and water, sequentially) and then dried under a nitrogen stream. The dried silanized surfaces were then heated in an oven at 110$^{\circ}$C for 20 minutes.   

\textit{Polymer brush growth:} In a typical procedure, the solvent solution (4:1 methanol:water, 50 mL) was first deoxygenated by bubbling with nitrogen for at least 30 minutes. SBMA (53.70 mmol, 15.0 g) was dissolved in 40 mL of the solvent solution to form the monomer solution, while bipy (1.344 mmol, 209.88 mg) and CuBr (0.5375 mmol, 77.11 mg) were dissolved in the remaining 10 mL of the solvent solution to form the catalytic solution. Both solutions were then stirred, while continuously being bubbled with nitrogen. After about 10 minutes, the catalytic solution was added to the monomer solution and was allowed to stir for another 2 minutes under nitrogen protection. The reaction mixture was then transferred to the reaction vessel containing the silanized surfaces. The reaction was performed under nitrogen protection. The reaction time was varied to achieve various brush heights. Upon completion of the polymer brush growth, the surfaces were rinsed with copious amounts of warm DI water (60$^{\circ}$C) and dried under a nitrogen stream. \\

\textbf{Boehmite surface preparation. }2 nm of Cr film was coated onto glass slides, followed by deposition of Al film up to 100 nm with a thermal evaporator system (Syskey, Taiwan) operating under high vacuum of $10^{-6}$--$10^{-7}$ Torr. Boiling in DI water close to 100$^{\circ}$C for 20 minutes converts the Al coating into nanostructured boehmite layer. \\  

\textbf{Patterning thiols on gold surfaces.} We first prepared 2 mM ethanol solutions of undecanolthiol (A) and sulfobetaine-3-undecanethiol (B). Using a cotton Q-tip, we swabbed the patterned side of the elastomeric stamp with solution A and dried the stamp in a stream of nitrogen gas for 30s. We then brought the stamp into contact with the gold substrate for 10 s, followed by immersing the gold substrate in solution B for 3 minutes. We then rinsed the patterned gold substrate with copious amounts of ethanol to remove any unadsorbed thiols. Once dried (in a stream of nitrogen gas), the patterned thiol surface is ready to be used. Detailed microcontact printing protocol can be found in \citeauthor{qin2010soft} (2010) \cite{qin2010soft}. \\     

\begin{figure}[!htb]
\centering
\includegraphics[scale=1.055]{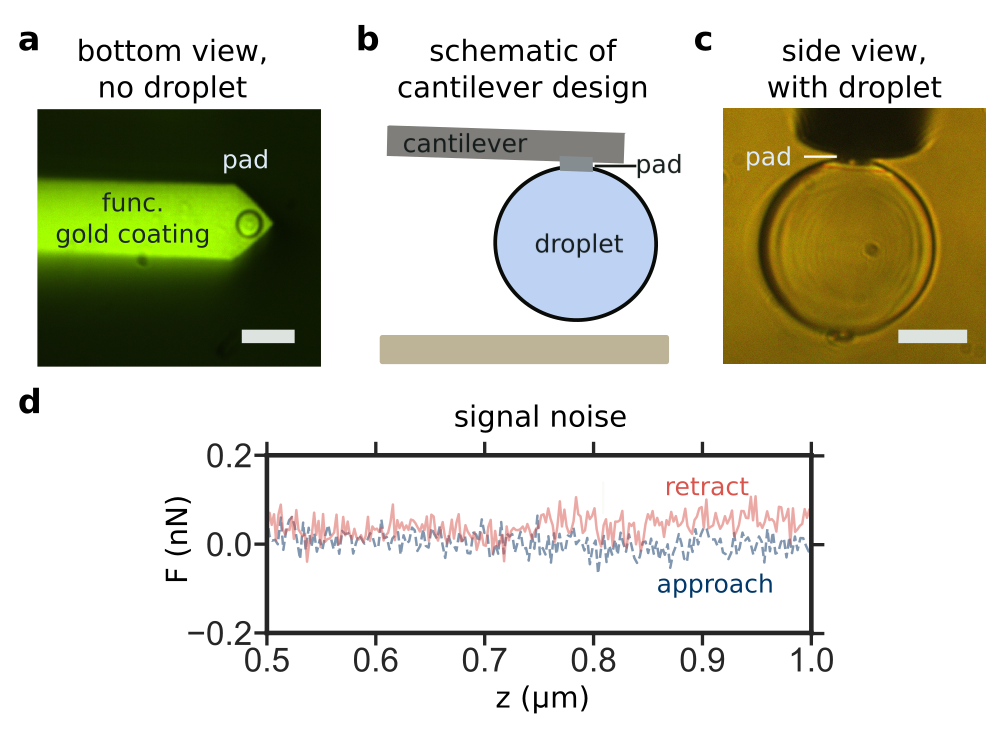}
\caption{\label{fig:cantilever} (a) Bottom-view of the rectangular cantilever with a 10-$\mu$m contact pad. (b, c) Our cantilever design allows us to confine droplet contact to the 10-$\mu$m pad. Scale bars are 20 $\mu$m. (d) Signal noise in retract and approach curves far away from the surface.}
\end{figure}

\textbf{Droplet probe AFM.} In our experiment, we used gold-coated (overall) tipless, rectangular cantilevers with spring constants $k_z$ = 0.2--2 N m$^{-1}$ (CSC37 series, MikroMash, dimensions: 350 $\mu$m length $\times$ 35 $\mu$m wide $\times$ 2 $\mu$m thick). Using direct-write, two-photon lithography technique (Nanoscribe$\textregistered$ Photonic Professional Instrument, Nanoscribe Inc., Germany), we created circular contact pads (10 $\mu$m diameter and 1 $\mu$m thick) on the cantilever tips (Fig.~\ref{fig:cantilever}a) \cite{goring2016tailored}. We then rendered the cantilever (except the contact pad) hydrophilic by immersing the cantilever in 2 mM ethanol solutions of sulfobetaine-3-undecanethiol for 1 min. 

To create small oil droplets, we force 50 $\mu$l of oil through a small capillary tube with inner and outer diameters of 360 $\mu$m and 290 $\mu$m into a petri-dish of water. This generates multiple droplets with a broad range of sizes $2R$ = 20---200 $\mu$m. We can then pick up a droplet of the desired size to perform force spectroscopy measurements on the surface of interest. Our cantilever design allows us to confine droplet contact to the 10-$\mu$m pad (Fig.~\ref{fig:cantilever}b, c). Note that while the contact pads improve the reproducibility of our results, they are not crucial and the experiments discussed in our paper can be performed with suitably functionalized tipless cantilevers without the contact pad \cite{daniel2019mapping}. 

Details on how we calibrated the cantilevers, performed the force spectroscopy measurements, and converted the raw voltage signals on the four-quadrant sensor into force values can be found in our previous publication \cite{daniel2019mapping}. 

There are many factors that limit the force sensitivity of an AFM \cite{smith1995limits}. Experimentally, we estimated $\Delta F$ to be about 20 pN, which we found by looking at the force fluctuations of the retract and approach curves far away from the surface (Fig.~\ref{fig:cantilever}d). 

\section*{Acknowledgements}

The authors are grateful to the Agency for Science, Technology and Research (A*STAR) for providing financial support under the SERC Career Development Award (grant number A1820g0089, project number SC25/18-8R111) and Pharos Advanced Surfaces Programme (grant number 1523700101, project number SC25/16-2P1203).

\bibliography{biblio}

\providecommand{\noopsort}[1]{}\providecommand{\singleletter}[1]{#1}%
\begin{thebibliography}{34}%
\makeatletter
\providecommand \@ifxundefined [1]{%
 \@ifx{#1\undefined}
}%
\providecommand \@ifnum [1]{%
 \ifnum #1\expandafter \@firstoftwo
 \else \expandafter \@secondoftwo
 \fi
}%
\providecommand \@ifx [1]{%
 \ifx #1\expandafter \@firstoftwo
 \else \expandafter \@secondoftwo
 \fi
}%
\providecommand \natexlab [1]{#1}%
\providecommand \enquote  [1]{``#1''}%
\providecommand \bibnamefont  [1]{#1}%
\providecommand \bibfnamefont [1]{#1}%
\providecommand \citenamefont [1]{#1}%
\providecommand \href@noop [0]{\@secondoftwo}%
\providecommand \href [0]{\begingroup \@sanitize@url \@href}%
\providecommand \@href[1]{\@@startlink{#1}\@@href}%
\providecommand \@@href[1]{\endgroup#1\@@endlink}%
\providecommand \@sanitize@url [0]{\catcode `\\12\catcode `\$12\catcode
  `\&12\catcode `\#12\catcode `\^12\catcode `\_12\catcode `\%12\relax}%
\providecommand \@@startlink[1]{}%
\providecommand \@@endlink[0]{}%
\providecommand \url  [0]{\begingroup\@sanitize@url \@url }%
\providecommand \@url [1]{\endgroup\@href {#1}{\urlprefix }}%
\providecommand \urlprefix  [0]{URL }%
\providecommand \Eprint [0]{\href }%
\providecommand \doibase [0]{http://dx.doi.org/}%
\providecommand \selectlanguage [0]{\@gobble}%
\providecommand \bibinfo  [0]{\@secondoftwo}%
\providecommand \bibfield  [0]{\@secondoftwo}%
\providecommand \translation [1]{[#1]}%
\providecommand \BibitemOpen [0]{}%
\providecommand \bibitemStop [0]{}%
\providecommand \bibitemNoStop [0]{.\EOS\space}%
\providecommand \EOS [0]{\spacefactor3000\relax}%
\providecommand \BibitemShut  [1]{\csname bibitem#1\endcsname}%
\let\auto@bib@innerbib\@empty
%</preamble>
\bibitem [{\citenamefont {Qu{\'e}r{\'e}}(2008)}]{quere2008wetting}%
  \BibitemOpen
  \bibfield  {author} {\bibinfo {author} {\bibfnamefont {D.}~\bibnamefont
  {Qu{\'e}r{\'e}}},\ }\bibfield  {title} {\enquote {\bibinfo {title} {Wetting
  and roughness},}\ }\href@noop {} {\bibfield  {journal} {\bibinfo  {journal}
  {Annu. Rev. Mater. Res.}\ }\textbf {\bibinfo {volume} {38}},\ \bibinfo
  {pages} {71--99} (\bibinfo {year} {2008})}\BibitemShut {NoStop}%
\bibitem [{\citenamefont {Bocquet}\ and\ \citenamefont
  {Lauga}(2011)}]{bocquet2011smooth}%
  \BibitemOpen
  \bibfield  {author} {\bibinfo {author} {\bibfnamefont {L.}~\bibnamefont
  {Bocquet}}\ and\ \bibinfo {author} {\bibfnamefont {E.}~\bibnamefont
  {Lauga}},\ }\bibfield  {title} {\enquote {\bibinfo {title} {A smooth
  future?}}\ }\href@noop {} {\bibfield  {journal} {\bibinfo  {journal} {Nat.
  Mater.}\ }\textbf {\bibinfo {volume} {10}},\ \bibinfo {pages} {334} (\bibinfo
  {year} {2011})}\BibitemShut {NoStop}%
\bibitem [{\citenamefont {Attinger}\ \emph {et~al.}(2014)\citenamefont
  {Attinger}, \citenamefont {Frankiewicz}, \citenamefont {Betz}, \citenamefont
  {Schutzius}, \citenamefont {Ganguly}, \citenamefont {Das}, \citenamefont
  {Kim},\ and\ \citenamefont {Megaridis}}]{attinger2014surface}%
  \BibitemOpen
  \bibfield  {author} {\bibinfo {author} {\bibfnamefont {D.}~\bibnamefont
  {Attinger}}, \bibinfo {author} {\bibfnamefont {C.}~\bibnamefont
  {Frankiewicz}}, \bibinfo {author} {\bibfnamefont {A.~R.}\ \bibnamefont
  {Betz}}, \bibinfo {author} {\bibfnamefont {T.~M.}\ \bibnamefont {Schutzius}},
  \bibinfo {author} {\bibfnamefont {R.}~\bibnamefont {Ganguly}}, \bibinfo
  {author} {\bibfnamefont {A.}~\bibnamefont {Das}}, \bibinfo {author}
  {\bibfnamefont {C.-J.}\ \bibnamefont {Kim}}, \ and\ \bibinfo {author}
  {\bibfnamefont {C.~M.}\ \bibnamefont {Megaridis}},\ }\bibfield  {title}
  {\enquote {\bibinfo {title} {Surface engineering for phase change heat
  transfer: A review},}\ }\href@noop {} {\bibfield  {journal} {\bibinfo
  {journal} {MRS Energy \& Sustainability}\ }\textbf {\bibinfo {volume} {1}}
  (\bibinfo {year} {2014})}\BibitemShut {NoStop}%
\bibitem [{\citenamefont {Liu}\ \emph {et~al.}(2019)\citenamefont {Liu},
  \citenamefont {Vuckovac}, \citenamefont {Latikka}, \citenamefont
  {Huhtam{\"a}ki},\ and\ \citenamefont {Ras}}]{Liu1147}%
  \BibitemOpen
  \bibfield  {author} {\bibinfo {author} {\bibfnamefont {K.}~\bibnamefont
  {Liu}}, \bibinfo {author} {\bibfnamefont {M.}~\bibnamefont {Vuckovac}},
  \bibinfo {author} {\bibfnamefont {M.}~\bibnamefont {Latikka}}, \bibinfo
  {author} {\bibfnamefont {T.}~\bibnamefont {Huhtam{\"a}ki}}, \ and\ \bibinfo
  {author} {\bibfnamefont {R.~H.~A.}\ \bibnamefont {Ras}},\ }\bibfield  {title}
  {\enquote {\bibinfo {title} {Improving surface-wetting characterization},}\
  }\href@noop {} {\bibfield  {journal} {\bibinfo  {journal} {Science}\ }\textbf
  {\bibinfo {volume} {363}},\ \bibinfo {pages} {1147--1148} (\bibinfo {year}
  {2019})}\BibitemShut {NoStop}%
\bibitem [{\citenamefont {Huhtam{\"a}ki}\ \emph {et~al.}(2018)\citenamefont
  {Huhtam{\"a}ki}, \citenamefont {Tian}, \citenamefont {Korhonen},\ and\
  \citenamefont {Ras}}]{huhtamaki2018surface}%
  \BibitemOpen
  \bibfield  {author} {\bibinfo {author} {\bibfnamefont {T.}~\bibnamefont
  {Huhtam{\"a}ki}}, \bibinfo {author} {\bibfnamefont {X.}~\bibnamefont {Tian}},
  \bibinfo {author} {\bibfnamefont {J.~T.}\ \bibnamefont {Korhonen}}, \ and\
  \bibinfo {author} {\bibfnamefont {R.~H.~A.}\ \bibnamefont {Ras}},\ }\bibfield
   {title} {\enquote {\bibinfo {title} {Surface-wetting characterization using
  contact-angle measurements},}\ }\href@noop {} {\bibfield  {journal} {\bibinfo
   {journal} {Nat. Protoc.}\ }\textbf {\bibinfo {volume} {13}},\ \bibinfo
  {pages} {1521} (\bibinfo {year} {2018})}\BibitemShut {NoStop}%
\bibitem [{\citenamefont {Samuel}\ \emph {et~al.}(2011)\citenamefont {Samuel},
  \citenamefont {Zhao},\ and\ \citenamefont {Law}}]{samuel2011study}%
  \BibitemOpen
  \bibfield  {author} {\bibinfo {author} {\bibfnamefont {B.}~\bibnamefont
  {Samuel}}, \bibinfo {author} {\bibfnamefont {H.}~\bibnamefont {Zhao}}, \ and\
  \bibinfo {author} {\bibfnamefont {K.-Y.}\ \bibnamefont {Law}},\ }\bibfield
  {title} {\enquote {\bibinfo {title} {Study of wetting and adhesion
  interactions between water and various polymer and superhydrophobic
  surfaces},}\ }\href@noop {} {\bibfield  {journal} {\bibinfo  {journal} {J.
  Phys. Chem. C}\ }\textbf {\bibinfo {volume} {115}},\ \bibinfo {pages}
  {14852--14861} (\bibinfo {year} {2011})}\BibitemShut {NoStop}%
\bibitem [{\citenamefont {Furmidge}(1962)}]{furmidge1962studies}%
  \BibitemOpen
  \bibfield  {author} {\bibinfo {author} {\bibfnamefont {C.~G.~L.}\
  \bibnamefont {Furmidge}},\ }\bibfield  {title} {\enquote {\bibinfo {title}
  {Studies at phase interfaces. i. the sliding of liquid drops on solid
  surfaces and a theory for spray retention},}\ }\href@noop {} {\bibfield
  {journal} {\bibinfo  {journal} {J. Colloid Sci.}\ }\textbf {\bibinfo {volume}
  {17}},\ \bibinfo {pages} {309--324} (\bibinfo {year} {1962})}\BibitemShut
  {NoStop}%
\bibitem [{\citenamefont {Decker}\ \emph {et~al.}(1999)\citenamefont {Decker},
  \citenamefont {Frank}, \citenamefont {Suo},\ and\ \citenamefont
  {Garoff}}]{decker1999physics}%
  \BibitemOpen
  \bibfield  {author} {\bibinfo {author} {\bibfnamefont {E.~L.}\ \bibnamefont
  {Decker}}, \bibinfo {author} {\bibfnamefont {B.}~\bibnamefont {Frank}},
  \bibinfo {author} {\bibfnamefont {Y.}~\bibnamefont {Suo}}, \ and\ \bibinfo
  {author} {\bibfnamefont {S.}~\bibnamefont {Garoff}},\ }\bibfield  {title}
  {\enquote {\bibinfo {title} {Physics of contact angle measurement},}\
  }\href@noop {} {\bibfield  {journal} {\bibinfo  {journal} {Colloids Surf. A
  Physicochem. Eng. Asp.}\ }\textbf {\bibinfo {volume} {156}},\ \bibinfo
  {pages} {177--189} (\bibinfo {year} {1999})}\BibitemShut {NoStop}%
\bibitem [{\citenamefont {Schellenberger}\ \emph {et~al.}(2016)\citenamefont
  {Schellenberger}, \citenamefont {Encinas}, \citenamefont {Vollmer},\ and\
  \citenamefont {Butt}}]{schellenberger2016water}%
  \BibitemOpen
  \bibfield  {author} {\bibinfo {author} {\bibfnamefont {F.}~\bibnamefont
  {Schellenberger}}, \bibinfo {author} {\bibfnamefont {N.}~\bibnamefont
  {Encinas}}, \bibinfo {author} {\bibfnamefont {D.}~\bibnamefont {Vollmer}}, \
  and\ \bibinfo {author} {\bibfnamefont {H.-J.}\ \bibnamefont {Butt}},\
  }\bibfield  {title} {\enquote {\bibinfo {title} {How water advances on
  superhydrophobic surfaces},}\ }\href@noop {} {\bibfield  {journal} {\bibinfo
  {journal} {Phys. Rev. Lett.}\ }\textbf {\bibinfo {volume} {116}},\ \bibinfo
  {pages} {096101} (\bibinfo {year} {2016})}\BibitemShut {NoStop}%
\bibitem [{\citenamefont {Daniel}\ \emph {et~al.}(2018)\citenamefont {Daniel},
  \citenamefont {Timonen}, \citenamefont {Li}, \citenamefont {Velling},
  \citenamefont {Kreder}, \citenamefont {Tetreault},\ and\ \citenamefont
  {Aizenberg}}]{daniel2018origins}%
  \BibitemOpen
  \bibfield  {author} {\bibinfo {author} {\bibfnamefont {D.}~\bibnamefont
  {Daniel}}, \bibinfo {author} {\bibfnamefont {J.~V.~I.}\ \bibnamefont
  {Timonen}}, \bibinfo {author} {\bibfnamefont {R.}~\bibnamefont {Li}},
  \bibinfo {author} {\bibfnamefont {S.~J.}\ \bibnamefont {Velling}}, \bibinfo
  {author} {\bibfnamefont {M.~J.}\ \bibnamefont {Kreder}}, \bibinfo {author}
  {\bibfnamefont {A.}~\bibnamefont {Tetreault}}, \ and\ \bibinfo {author}
  {\bibfnamefont {J.}~\bibnamefont {Aizenberg}},\ }\bibfield  {title} {\enquote
  {\bibinfo {title} {Origins of extreme liquid repellency on structured, flat,
  and lubricated hydrophobic surfaces},}\ }\href@noop {} {\bibfield  {journal}
  {\bibinfo  {journal} {Phys. Rev. Lett.}\ }\textbf {\bibinfo {volume} {120}},\
  \bibinfo {pages} {244503} (\bibinfo {year} {2018})}\BibitemShut {NoStop}%
\bibitem [{\citenamefont {Liimatainen}\ \emph {et~al.}(2017)\citenamefont
  {Liimatainen}, \citenamefont {Vuckovac}, \citenamefont {Jokinen},
  \citenamefont {Sariola}, \citenamefont {Hokkanen}, \citenamefont {Zhou},\
  and\ \citenamefont {Ras}}]{liimatainen2017mapping}%
  \BibitemOpen
  \bibfield  {author} {\bibinfo {author} {\bibfnamefont {V.}~\bibnamefont
  {Liimatainen}}, \bibinfo {author} {\bibfnamefont {M.}~\bibnamefont
  {Vuckovac}}, \bibinfo {author} {\bibfnamefont {V.}~\bibnamefont {Jokinen}},
  \bibinfo {author} {\bibfnamefont {V.}~\bibnamefont {Sariola}}, \bibinfo
  {author} {\bibfnamefont {M.~J.}\ \bibnamefont {Hokkanen}}, \bibinfo {author}
  {\bibfnamefont {Q.}~\bibnamefont {Zhou}}, \ and\ \bibinfo {author}
  {\bibfnamefont {R.~H.~A.}\ \bibnamefont {Ras}},\ }\bibfield  {title}
  {\enquote {\bibinfo {title} {Mapping microscale wetting variations on
  biological and synthetic water-repellent surfaces},}\ }\href@noop {}
  {\bibfield  {journal} {\bibinfo  {journal} {Nat. Comm.}\ }\textbf {\bibinfo
  {volume} {8}},\ \bibinfo {pages} {1798} (\bibinfo {year} {2017})}\BibitemShut
  {NoStop}%
\bibitem [{\citenamefont {Srinivasan}\ \emph {et~al.}(2011)\citenamefont
  {Srinivasan}, \citenamefont {McKinley},\ and\ \citenamefont
  {Cohen}}]{srinivasan2011assessing}%
  \BibitemOpen
  \bibfield  {author} {\bibinfo {author} {\bibfnamefont {S.}~\bibnamefont
  {Srinivasan}}, \bibinfo {author} {\bibfnamefont {G.~H.}\ \bibnamefont
  {McKinley}}, \ and\ \bibinfo {author} {\bibfnamefont {R.~E.}\ \bibnamefont
  {Cohen}},\ }\bibfield  {title} {\enquote {\bibinfo {title} {Assessing the
  accuracy of contact angle measurements for sessile drops on liquid-repellent
  surfaces},}\ }\href@noop {} {\bibfield  {journal} {\bibinfo  {journal}
  {Langmuir}\ }\textbf {\bibinfo {volume} {27}},\ \bibinfo {pages}
  {13582--13589} (\bibinfo {year} {2011})}\BibitemShut {NoStop}%
\bibitem [{\citenamefont {Tadmor}\ \emph {et~al.}(2009)\citenamefont {Tadmor},
  \citenamefont {Bahadur}, \citenamefont {Leh}, \citenamefont {N’guessan},
  \citenamefont {Jaini},\ and\ \citenamefont {Dang}}]{tadmor2009measurement}%
  \BibitemOpen
  \bibfield  {author} {\bibinfo {author} {\bibfnamefont {R.}~\bibnamefont
  {Tadmor}}, \bibinfo {author} {\bibfnamefont {P.}~\bibnamefont {Bahadur}},
  \bibinfo {author} {\bibfnamefont {A.}~\bibnamefont {Leh}}, \bibinfo {author}
  {\bibfnamefont {H.~E.}\ \bibnamefont {N’guessan}}, \bibinfo {author}
  {\bibfnamefont {R.}~\bibnamefont {Jaini}}, \ and\ \bibinfo {author}
  {\bibfnamefont {L.}~\bibnamefont {Dang}},\ }\bibfield  {title} {\enquote
  {\bibinfo {title} {Measurement of lateral adhesion forces at the interface
  between a liquid drop and a substrate},}\ }\href@noop {} {\bibfield
  {journal} {\bibinfo  {journal} {Phys. Rev. Lett.}\ }\textbf {\bibinfo
  {volume} {103}},\ \bibinfo {pages} {266101} (\bibinfo {year}
  {2009})}\BibitemShut {NoStop}%
\bibitem [{\citenamefont {Daniel}\ \emph {et~al.}(2017)\citenamefont {Daniel},
  \citenamefont {Timonen}, \citenamefont {Li}, \citenamefont {Velling},\ and\
  \citenamefont {Aizenberg}}]{daniel2017oleoplaning}%
  \BibitemOpen
  \bibfield  {author} {\bibinfo {author} {\bibfnamefont {D.}~\bibnamefont
  {Daniel}}, \bibinfo {author} {\bibfnamefont {J.~V.~I.}\ \bibnamefont
  {Timonen}}, \bibinfo {author} {\bibfnamefont {R.}~\bibnamefont {Li}},
  \bibinfo {author} {\bibfnamefont {S.~J.}\ \bibnamefont {Velling}}, \ and\
  \bibinfo {author} {\bibfnamefont {J.}~\bibnamefont {Aizenberg}},\ }\bibfield
  {title} {\enquote {\bibinfo {title} {Oleoplaning droplets on lubricated
  surfaces},}\ }\href@noop {} {\bibfield  {journal} {\bibinfo  {journal} {Nat.
  Phys.}\ }\textbf {\bibinfo {volume} {13}},\ \bibinfo {pages} {1020} (\bibinfo
  {year} {2017})}\BibitemShut {NoStop}%
\bibitem [{\citenamefont {Timonen}\ \emph {et~al.}(2013)\citenamefont
  {Timonen}, \citenamefont {Latikka}, \citenamefont {Ikkala},\ and\
  \citenamefont {Ras}}]{timonen2013free}%
  \BibitemOpen
  \bibfield  {author} {\bibinfo {author} {\bibfnamefont {J.~V.~I.}\
  \bibnamefont {Timonen}}, \bibinfo {author} {\bibfnamefont {M.}~\bibnamefont
  {Latikka}}, \bibinfo {author} {\bibfnamefont {O.}~\bibnamefont {Ikkala}}, \
  and\ \bibinfo {author} {\bibfnamefont {R.~H.~A}\ \bibnamefont {Ras}},\
  }\bibfield  {title} {\enquote {\bibinfo {title} {Free-decay and resonant
  methods for investigating the fundamental limit of superhydrophobicity},}\
  }\href@noop {} {\bibfield  {journal} {\bibinfo  {journal} {Nat. Comm.}\
  }\textbf {\bibinfo {volume} {4}},\ \bibinfo {pages} {2398} (\bibinfo {year}
  {2013})}\BibitemShut {NoStop}%
\bibitem [{\citenamefont {Daniel}\ \emph
  {et~al.}(2019{\natexlab{a}})\citenamefont {Daniel}, \citenamefont {Lay},
  \citenamefont {Sng}, \citenamefont {Lee}, \citenamefont {Neo}, \citenamefont
  {Ling},\ and\ \citenamefont {Tomczak}}]{daniel2019mapping}%
  \BibitemOpen
  \bibfield  {author} {\bibinfo {author} {\bibfnamefont {D.}~\bibnamefont
  {Daniel}}, \bibinfo {author} {\bibfnamefont {C.~L.}\ \bibnamefont {Lay}},
  \bibinfo {author} {\bibfnamefont {A.}~\bibnamefont {Sng}}, \bibinfo {author}
  {\bibfnamefont {C.~J.~J.}\ \bibnamefont {Lee}}, \bibinfo {author}
  {\bibfnamefont {D.~C.~J.}\ \bibnamefont {Neo}}, \bibinfo {author}
  {\bibfnamefont {X.~Y.}\ \bibnamefont {Ling}}, \ and\ \bibinfo {author}
  {\bibfnamefont {N.}~\bibnamefont {Tomczak}},\ }\bibfield  {title} {\enquote
  {\bibinfo {title} {Mapping micrometer-scale wetting properties of
  superhydrophobic surfaces},}\ }\href@noop {} {\bibfield  {journal} {\bibinfo
  {journal} {Proc. Natl. Acad. Sci. U.S.A.}\ } (\bibinfo {year}
  {2019}{\natexlab{a}})}\BibitemShut {NoStop}%
\bibitem [{\citenamefont {Xie}\ \emph {et~al.}(2017)\citenamefont {Xie},
  \citenamefont {Shi}, \citenamefont {Cui},\ and\ \citenamefont
  {Zeng}}]{xie2017surface}%
  \BibitemOpen
  \bibfield  {author} {\bibinfo {author} {\bibfnamefont {L.}~\bibnamefont
  {Xie}}, \bibinfo {author} {\bibfnamefont {C.}~\bibnamefont {Shi}}, \bibinfo
  {author} {\bibfnamefont {X.}~\bibnamefont {Cui}}, \ and\ \bibinfo {author}
  {\bibfnamefont {H.}~\bibnamefont {Zeng}},\ }\bibfield  {title} {\enquote
  {\bibinfo {title} {Surface forces and interaction mechanisms of emulsion
  drops and gas bubbles in complex fluids},}\ }\href@noop {} {\bibfield
  {journal} {\bibinfo  {journal} {Langmuir}\ }\textbf {\bibinfo {volume}
  {33}},\ \bibinfo {pages} {3911--3925} (\bibinfo {year} {2017})}\BibitemShut
  {NoStop}%
\bibitem [{\citenamefont {Shi}\ \emph {et~al.}(2016)\citenamefont {Shi},
  \citenamefont {Yan}, \citenamefont {Xie}, \citenamefont {Zhang},
  \citenamefont {Wang}, \citenamefont {Takahara},\ and\ \citenamefont
  {Zeng}}]{shi2016long}%
  \BibitemOpen
  \bibfield  {author} {\bibinfo {author} {\bibfnamefont {C.}~\bibnamefont
  {Shi}}, \bibinfo {author} {\bibfnamefont {B.}~\bibnamefont {Yan}}, \bibinfo
  {author} {\bibfnamefont {L.}~\bibnamefont {Xie}}, \bibinfo {author}
  {\bibfnamefont {L.}~\bibnamefont {Zhang}}, \bibinfo {author} {\bibfnamefont
  {J.}~\bibnamefont {Wang}}, \bibinfo {author} {\bibfnamefont {A.}~\bibnamefont
  {Takahara}}, \ and\ \bibinfo {author} {\bibfnamefont {H.}~\bibnamefont
  {Zeng}},\ }\bibfield  {title} {\enquote {\bibinfo {title} {Long-range
  hydrophilic attraction between water and polyelectrolyte surfaces in oil},}\
  }\href@noop {} {\bibfield  {journal} {\bibinfo  {journal} {Angew. Chem. Int.
  Ed.}\ }\textbf {\bibinfo {volume} {55}},\ \bibinfo {pages} {15017--15021}
  (\bibinfo {year} {2016})}\BibitemShut {NoStop}%
\bibitem [{\citenamefont {Manor}\ \emph {et~al.}(2008)\citenamefont {Manor},
  \citenamefont {Vakarelski}, \citenamefont {Tang}, \citenamefont {{O'Shea}},
  \citenamefont {Stevens}, \citenamefont {Grieser}, \citenamefont {Dagastine},\
  and\ \citenamefont {Chan}}]{manor2008hydrodynamic}%
  \BibitemOpen
  \bibfield  {author} {\bibinfo {author} {\bibfnamefont {O.}~\bibnamefont
  {Manor}}, \bibinfo {author} {\bibfnamefont {I.~U.}\ \bibnamefont
  {Vakarelski}}, \bibinfo {author} {\bibfnamefont {X.}~\bibnamefont {Tang}},
  \bibinfo {author} {\bibfnamefont {S.~J.}\ \bibnamefont {{O'Shea}}}, \bibinfo
  {author} {\bibfnamefont {G.~W.}\ \bibnamefont {Stevens}}, \bibinfo {author}
  {\bibfnamefont {F.}~\bibnamefont {Grieser}}, \bibinfo {author} {\bibfnamefont
  {R.~R.}\ \bibnamefont {Dagastine}}, \ and\ \bibinfo {author} {\bibfnamefont
  {D.~Y.~C.}\ \bibnamefont {Chan}},\ }\bibfield  {title} {\enquote {\bibinfo
  {title} {Hydrodynamic boundary conditions and dynamic forces between bubbles
  and surfaces},}\ }\href@noop {} {\bibfield  {journal} {\bibinfo  {journal}
  {Phys. Rev. Lett.}\ }\textbf {\bibinfo {volume} {101}},\ \bibinfo {pages}
  {024501} (\bibinfo {year} {2008})}\BibitemShut {NoStop}%
\bibitem [{\citenamefont {Escobar}\ \emph {et~al.}(2017)\citenamefont
  {Escobar}, \citenamefont {Garza},\ and\ \citenamefont
  {Castillo}}]{escobar2017measuring}%
  \BibitemOpen
  \bibfield  {author} {\bibinfo {author} {\bibfnamefont {J.~V.}\ \bibnamefont
  {Escobar}}, \bibinfo {author} {\bibfnamefont {C.}~\bibnamefont {Garza}}, \
  and\ \bibinfo {author} {\bibfnamefont {R.}~\bibnamefont {Castillo}},\
  }\bibfield  {title} {\enquote {\bibinfo {title} {Measuring adhesion on rough
  surfaces using atomic force microscopy with a liquid probe},}\ }\href@noop {}
  {\bibfield  {journal} {\bibinfo  {journal} {Beilstein J. Nanotechnol.}\
  }\textbf {\bibinfo {volume} {8}},\ \bibinfo {pages} {813--825} (\bibinfo
  {year} {2017})}\BibitemShut {NoStop}%
\bibitem [{\citenamefont {Daniel}\ \emph
  {et~al.}(2019{\natexlab{b}})\citenamefont {Daniel}, \citenamefont {Chia},
  \citenamefont {Moh}, \citenamefont {Liu}, \citenamefont {Koh}, \citenamefont
  {Zhang},\ and\ \citenamefont {Tomczak}}]{daniel2019hydration}%
  \BibitemOpen
  \bibfield  {author} {\bibinfo {author} {\bibfnamefont {D.}~\bibnamefont
  {Daniel}}, \bibinfo {author} {\bibfnamefont {A.~Y.~T.}\ \bibnamefont {Chia}},
  \bibinfo {author} {\bibfnamefont {L.~C.~H.}\ \bibnamefont {Moh}}, \bibinfo
  {author} {\bibfnamefont {R.}~\bibnamefont {Liu}}, \bibinfo {author}
  {\bibfnamefont {X.~Q.}\ \bibnamefont {Koh}}, \bibinfo {author} {\bibfnamefont
  {X.}~\bibnamefont {Zhang}}, \ and\ \bibinfo {author} {\bibfnamefont
  {N.}~\bibnamefont {Tomczak}},\ }\bibfield  {title} {\enquote {\bibinfo
  {title} {Hydration lubrication of polyzwitterionic brushes leads to nearly
  friction- and adhesion-free droplet motion},}\ }\href@noop {} {\bibfield
  {journal} {\bibinfo  {journal} {Comm. Phys.}\ } (\bibinfo {year}
  {2019}{\natexlab{b}})}\BibitemShut {NoStop}%
\bibitem [{\citenamefont {Azzaroni}\ \emph {et~al.}(2006)\citenamefont
  {Azzaroni}, \citenamefont {Brown},\ and\ \citenamefont
  {Huck}}]{azzaroni2006ucst}%
  \BibitemOpen
  \bibfield  {author} {\bibinfo {author} {\bibfnamefont {O.}~\bibnamefont
  {Azzaroni}}, \bibinfo {author} {\bibfnamefont {A.~A.}\ \bibnamefont {Brown}},
  \ and\ \bibinfo {author} {\bibfnamefont {W.~T.~S.}\ \bibnamefont {Huck}},\
  }\bibfield  {title} {\enquote {\bibinfo {title} {{UCST wetting transitions of
  polyzwitterionic brushes driven by self-association}},}\ }\href@noop {}
  {\bibfield  {journal} {\bibinfo  {journal} {Angew. Chem. Int. Ed.}\ }\textbf
  {\bibinfo {volume} {45}},\ \bibinfo {pages} {1770--1774} (\bibinfo {year}
  {2006})}\BibitemShut {NoStop}%
\bibitem [{\citenamefont {Kobayashi}\ \emph {et~al.}(2012)\citenamefont
  {Kobayashi}, \citenamefont {Terayama}, \citenamefont {Yamaguchi},
  \citenamefont {Terada}, \citenamefont {Murakami}, \citenamefont {Ishihara},\
  and\ \citenamefont {Takahara}}]{kobayashi2012wettability}%
  \BibitemOpen
  \bibfield  {author} {\bibinfo {author} {\bibfnamefont {M.}~\bibnamefont
  {Kobayashi}}, \bibinfo {author} {\bibfnamefont {Y.}~\bibnamefont {Terayama}},
  \bibinfo {author} {\bibfnamefont {H.}~\bibnamefont {Yamaguchi}}, \bibinfo
  {author} {\bibfnamefont {M.}~\bibnamefont {Terada}}, \bibinfo {author}
  {\bibfnamefont {D.}~\bibnamefont {Murakami}}, \bibinfo {author}
  {\bibfnamefont {K.}~\bibnamefont {Ishihara}}, \ and\ \bibinfo {author}
  {\bibfnamefont {A.}~\bibnamefont {Takahara}},\ }\bibfield  {title} {\enquote
  {\bibinfo {title} {Wettability and antifouling behavior on the surfaces of
  superhydrophilic polymer brushes},}\ }\href@noop {} {\bibfield  {journal}
  {\bibinfo  {journal} {Langmuir}\ }\textbf {\bibinfo {volume} {28}},\ \bibinfo
  {pages} {7212--7222} (\bibinfo {year} {2012})}\BibitemShut {NoStop}%
\bibitem [{\citenamefont {Liu}\ \emph {et~al.}(2017)\citenamefont {Liu},
  \citenamefont {Wang},\ and\ \citenamefont {Jiang}}]{liu2017nature}%
  \BibitemOpen
  \bibfield  {author} {\bibinfo {author} {\bibfnamefont {M.}~\bibnamefont
  {Liu}}, \bibinfo {author} {\bibfnamefont {S.}~\bibnamefont {Wang}}, \ and\
  \bibinfo {author} {\bibfnamefont {L.}~\bibnamefont {Jiang}},\ }\bibfield
  {title} {\enquote {\bibinfo {title} {Nature-inspired superwettability
  systems},}\ }\href@noop {} {\bibfield  {journal} {\bibinfo  {journal} {Nat.
  Rev. Mater.}\ }\textbf {\bibinfo {volume} {2}},\ \bibinfo {pages} {17036}
  (\bibinfo {year} {2017})}\BibitemShut {NoStop}%
\bibitem [{\citenamefont {Evans}(2001)}]{evans2001probing}%
  \BibitemOpen
  \bibfield  {author} {\bibinfo {author} {\bibfnamefont {E.}~\bibnamefont
  {Evans}},\ }\bibfield  {title} {\enquote {\bibinfo {title} {Probing the
  relation between force—lifetime—and chemistry in single molecular
  bonds},}\ }\href@noop {} {\bibfield  {journal} {\bibinfo  {journal} {Annu.
  Rev. Biophys. Biomol. Struct.}\ }\textbf {\bibinfo {volume} {30}},\ \bibinfo
  {pages} {105--128} (\bibinfo {year} {2001})}\BibitemShut {NoStop}%
\bibitem [{\citenamefont {Liu}\ \emph {et~al.}(2009)\citenamefont {Liu},
  \citenamefont {Wang}, \citenamefont {Wei}, \citenamefont {Song},\ and\
  \citenamefont {Jiang}}]{liu2009bioinspired}%
  \BibitemOpen
  \bibfield  {author} {\bibinfo {author} {\bibfnamefont {M.}~\bibnamefont
  {Liu}}, \bibinfo {author} {\bibfnamefont {S.}~\bibnamefont {Wang}}, \bibinfo
  {author} {\bibfnamefont {Z.}~\bibnamefont {Wei}}, \bibinfo {author}
  {\bibfnamefont {Y.}~\bibnamefont {Song}}, \ and\ \bibinfo {author}
  {\bibfnamefont {L.}~\bibnamefont {Jiang}},\ }\bibfield  {title} {\enquote
  {\bibinfo {title} {Bioinspired design of a superoleophobic and low adhesive
  water/solid interface},}\ }\href@noop {} {\bibfield  {journal} {\bibinfo
  {journal} {Adv. Mater.}\ }\textbf {\bibinfo {volume} {21}},\ \bibinfo {pages}
  {665--669} (\bibinfo {year} {2009})}\BibitemShut {NoStop}%
\bibitem [{\citenamefont {Qin}\ \emph {et~al.}(2010)\citenamefont {Qin},
  \citenamefont {Xia},\ and\ \citenamefont {Whitesides}}]{qin2010soft}%
  \BibitemOpen
  \bibfield  {author} {\bibinfo {author} {\bibfnamefont {D.}~\bibnamefont
  {Qin}}, \bibinfo {author} {\bibfnamefont {Y.}~\bibnamefont {Xia}}, \ and\
  \bibinfo {author} {\bibfnamefont {G.~M.}\ \bibnamefont {Whitesides}},\
  }\bibfield  {title} {\enquote {\bibinfo {title} {Soft lithography for
  micro-and nanoscale patterning},}\ }\href@noop {} {\bibfield  {journal}
  {\bibinfo  {journal} {Nat. Protoc.}\ }\textbf {\bibinfo {volume} {5}},\
  \bibinfo {pages} {491} (\bibinfo {year} {2010})}\BibitemShut {NoStop}%
\bibitem [{\citenamefont {Frisbie}\ \emph {et~al.}(1994)\citenamefont
  {Frisbie}, \citenamefont {Rozsnyai}, \citenamefont {Noy}, \citenamefont
  {Wrighton},\ and\ \citenamefont {Lieber}}]{frisbie1994functional}%
  \BibitemOpen
  \bibfield  {author} {\bibinfo {author} {\bibfnamefont {C.~D.}\ \bibnamefont
  {Frisbie}}, \bibinfo {author} {\bibfnamefont {L.~F.}\ \bibnamefont
  {Rozsnyai}}, \bibinfo {author} {\bibfnamefont {A.}~\bibnamefont {Noy}},
  \bibinfo {author} {\bibfnamefont {M.~S.}\ \bibnamefont {Wrighton}}, \ and\
  \bibinfo {author} {\bibfnamefont {C.~M.}\ \bibnamefont {Lieber}},\ }\bibfield
   {title} {\enquote {\bibinfo {title} {Functional group imaging by chemical
  force microscopy},}\ }\href@noop {} {\bibfield  {journal} {\bibinfo
  {journal} {Science}\ }\textbf {\bibinfo {volume} {265}},\ \bibinfo {pages}
  {2071--2074} (\bibinfo {year} {1994})}\BibitemShut {NoStop}%
\bibitem [{\citenamefont {Wong}\ \emph {et~al.}(1998)\citenamefont {Wong},
  \citenamefont {Joselevich}, \citenamefont {Woolley}, \citenamefont {Cheung},\
  and\ \citenamefont {Lieber}}]{wong1998covalently}%
  \BibitemOpen
  \bibfield  {author} {\bibinfo {author} {\bibfnamefont {S.~S.}\ \bibnamefont
  {Wong}}, \bibinfo {author} {\bibfnamefont {E.}~\bibnamefont {Joselevich}},
  \bibinfo {author} {\bibfnamefont {A.~T.}\ \bibnamefont {Woolley}}, \bibinfo
  {author} {\bibfnamefont {C.~L.}\ \bibnamefont {Cheung}}, \ and\ \bibinfo
  {author} {\bibfnamefont {C.~M.}\ \bibnamefont {Lieber}},\ }\bibfield  {title}
  {\enquote {\bibinfo {title} {Covalently functionalized nanotubes as
  nanometre-sized probes in chemistry and biology},}\ }\href@noop {} {\bibfield
   {journal} {\bibinfo  {journal} {Nature}\ }\textbf {\bibinfo {volume}
  {394}},\ \bibinfo {pages} {52} (\bibinfo {year} {1998})}\BibitemShut
  {NoStop}%
\bibitem [{\citenamefont {Schönherr}\ \emph {et~al.}(2005)\citenamefont
  {Schönherr}, \citenamefont {Feng}, \citenamefont {Tomczak},\ and\
  \citenamefont {Vancso}}]{schonherr2005}%
  \BibitemOpen
  \bibfield  {author} {\bibinfo {author} {\bibfnamefont {Holger}\ \bibnamefont
  {Schönherr}}, \bibinfo {author} {\bibfnamefont {Chuan~Liang}\ \bibnamefont
  {Feng}}, \bibinfo {author} {\bibfnamefont {Nikodem}\ \bibnamefont {Tomczak}},
  \ and\ \bibinfo {author} {\bibfnamefont {G.~Julius}\ \bibnamefont {Vancso}},\
  }\bibfield  {title} {\enquote {\bibinfo {title} {Compositional mapping of
  polymer surfaces by chemical force microscopy down to the nanometer scale:
  Reactions in block copolymer microdomains},}\ }\href@noop {} {\bibfield
  {journal} {\bibinfo  {journal} {Macromolecular Symposia}\ }\textbf {\bibinfo
  {volume} {230}},\ \bibinfo {pages} {149--157} (\bibinfo {year}
  {2005})}\BibitemShut {NoStop}%
\bibitem [{\citenamefont {Hoek}\ \emph {et~al.}(2003)\citenamefont {Hoek},
  \citenamefont {Bhattacharjee},\ and\ \citenamefont
  {Elimelech}}]{hoek2003effect}%
  \BibitemOpen
  \bibfield  {author} {\bibinfo {author} {\bibfnamefont {E.~M.~V.}\
  \bibnamefont {Hoek}}, \bibinfo {author} {\bibfnamefont {S.}~\bibnamefont
  {Bhattacharjee}}, \ and\ \bibinfo {author} {\bibfnamefont {M.}~\bibnamefont
  {Elimelech}},\ }\bibfield  {title} {\enquote {\bibinfo {title} {Effect of
  membrane surface roughness on colloid- membrane dlvo interactions},}\
  }\href@noop {} {\bibfield  {journal} {\bibinfo  {journal} {Langmuir}\
  }\textbf {\bibinfo {volume} {19}},\ \bibinfo {pages} {4836--4847} (\bibinfo
  {year} {2003})}\BibitemShut {NoStop}%
\bibitem [{\citenamefont {Mi}\ and\ \citenamefont
  {Elimelech}(2010)}]{mi2010organic}%
  \BibitemOpen
  \bibfield  {author} {\bibinfo {author} {\bibfnamefont {B.}~\bibnamefont
  {Mi}}\ and\ \bibinfo {author} {\bibfnamefont {M.}~\bibnamefont {Elimelech}},\
  }\bibfield  {title} {\enquote {\bibinfo {title} {Organic fouling of forward
  osmosis membranes: fouling reversibility and cleaning without chemical
  reagents},}\ }\href@noop {} {\bibfield  {journal} {\bibinfo  {journal} {J.
  Membr. Sci.}\ }\textbf {\bibinfo {volume} {348}},\ \bibinfo {pages}
  {337--345} (\bibinfo {year} {2010})}\BibitemShut {NoStop}%
\bibitem [{\citenamefont {G{\"o}ring}\ \emph {et~al.}(2016)\citenamefont
  {G{\"o}ring}, \citenamefont {Dietrich}, \citenamefont {Blaicher},
  \citenamefont {Sharma}, \citenamefont {Korvink}, \citenamefont {Schimmel},
  \citenamefont {Koos},\ and\ \citenamefont
  {H{\"o}lscher}}]{goring2016tailored}%
  \BibitemOpen
  \bibfield  {author} {\bibinfo {author} {\bibfnamefont {G.}~\bibnamefont
  {G{\"o}ring}}, \bibinfo {author} {\bibfnamefont {P.-I.}\ \bibnamefont
  {Dietrich}}, \bibinfo {author} {\bibfnamefont {M.}~\bibnamefont {Blaicher}},
  \bibinfo {author} {\bibfnamefont {S.}~\bibnamefont {Sharma}}, \bibinfo
  {author} {\bibfnamefont {J.~G.}\ \bibnamefont {Korvink}}, \bibinfo {author}
  {\bibfnamefont {T.}~\bibnamefont {Schimmel}}, \bibinfo {author}
  {\bibfnamefont {C.}~\bibnamefont {Koos}}, \ and\ \bibinfo {author}
  {\bibfnamefont {H.}~\bibnamefont {H{\"o}lscher}},\ }\bibfield  {title}
  {\enquote {\bibinfo {title} {Tailored probes for atomic force microscopy
  fabricated by two-photon polymerization},}\ }\href@noop {} {\bibfield
  {journal} {\bibinfo  {journal} {Appl. Phys. Lett.}\ }\textbf {\bibinfo
  {volume} {109}},\ \bibinfo {pages} {063101} (\bibinfo {year}
  {2016})}\BibitemShut {NoStop}%
\bibitem [{\citenamefont {Smith}(1995)}]{smith1995limits}%
  \BibitemOpen
  \bibfield  {author} {\bibinfo {author} {\bibfnamefont {D.~P.~E.}\
  \bibnamefont {Smith}},\ }\bibfield  {title} {\enquote {\bibinfo {title}
  {Limits of force microscopy},}\ }\href@noop {} {\bibfield  {journal}
  {\bibinfo  {journal} {Rev. Sci. Instrum.}\ }\textbf {\bibinfo {volume}
  {66}},\ \bibinfo {pages} {3191--3195} (\bibinfo {year} {1995})}\BibitemShut
  {NoStop}%
\end{thebibliography}%
\end{document}